\documentclass[journal,twocolumn]{IEEEtran}
\usepackage{}
\usepackage{graphicx}
\usepackage{epstopdf}
\usepackage{amssymb}
\usepackage{amsfonts}
\usepackage{amsmath}
\usepackage{algorithm}
\usepackage{algorithmic}
\usepackage{subeqnarray}
\usepackage{cases}
\usepackage{color}

\usepackage{bm}
\usepackage{subfigure,amsmath,amssymb,cite}
\usepackage{float}
\ifCLASSINFOpdf
\else
\fi
\begin{document}

\title{ Two Low-complexity DOA Estimators for Massive/Ultra-massive MIMO Receive Array}

\author{ Yiwen Chen, Xichao Zhan, Feng Shu, Qijuan Jie, Xin Cheng, \\
Zhihong Zhuang, and Jiangzhou Wang, \emph{ Fellow, IEEE}
\thanks{This work was supported in part by the National Natural Science Foundation of China (Nos. 62071234, 62071289, and 61972093), the Hainan Province Science and Technology Special Fund (ZDKJ2021022), the Scientific Research Fund Project of Hainan University under Grant KYQD(ZR)-21008, and the National Key R\&D Program of China under Grant 2018YFB180110. (Corresponding authors: Feng Shu and Zhihong Zhuang).}
\thanks{Yiwen Chen, Xichao Zhan, Qijuan Jie, and Feng Shu are with the School of Information and Communication Engineering, Hainan University, Haikou, 570228, China. (Email: shufeng0101@163.com).}
\thanks{Feng Shu, Xin Cheng, and Zhihong Zhuang are with the School of Electronic and Optical Engineering, Nanjing University of Science and Technology, 210094, CHINA. (Email: shufeng0101 @163.com).}
\thanks{Jiangzhou Wang is with the School of Engineering, University of Kent, Canterbury CT27NT, U.K. (Email: {j.z.wang}@kent.ac.uk).}
}
\maketitle

\begin{abstract}
Eigen-decomposition-based direction finding methods of using large-scale/ultra-large-scale fully-digital receive antenna arrays lead to a high or ultra-high complexity. To address the complexity dilemma, in this paper, three low-complexity estimators are proposed: partitioned subarray auto-correlation combining (PSAC), partitioned subarray cross-correlation combining (PSCC) and power iteration max correlation successive convex approximation (PI-Max-CSCA). Compared with the conventional no-partitioned direction finding method like root multiple signal classification (Root-MUSIC), in the PSAC method, the total set of antennas are equally partitioned into subsets of antennas, called subarrays, each subarray performs independent DOA estimation, and all DOA estimates are coherently combined to give the final estimation. For a better performance, the cross-correlation among sub-arrays is further exploited in the PSCC method to achieve the near-Cramer-Rao lower bound (CRLB) performance with the help of auto-correlation. To further reduce the complexity, in the PI-Max-CSCA method, using a fraction of all subarrays to make an initial coarse direction measurement (ICDM), the power iterative method is adopted to compute the more precise steering vector (SV) by exploiting the total array, and a more accurate DOA value is found using ICDM and SV through the maximum correlation method solved by successive convex approximation.
Simulation results show that as the number of antennas goes to large-scale, the proposed three methods can achieve a dramatic complexity reduction over conventional Root-MUISC. Particularly, the PSCC and PI-Max-CSCA can reach the CRLB while the PSAC shows a substantial performance loss.
\end{abstract}
\section{Introduction}
Wireless direction of arrival (DOA) technology is a crucial problem in wireless networks, and radar. It has been widely used in various modern engineering fields, including navigation, unmanned aerial vehicle (UAV) communication, intelligent transportation, millimeter-wave (mmWave) communications \cite{7400949} and so on.
Before performing DOA measurement, it is key to infer whether the emitters exist or not. If there is no emitter, it is apparent that there is no demand to do DOA estimation. Only and only if emitters exist, the next step of DOA measurement is required. In \cite{9695432}, the authors have proposed three high-performance detectors to infer the existence of passive emitters from the eigen-space of the sample covariance matrix of the received signal. Furthermore, several machine learning methods of detecting the number of passive emitters were developed to improve the accuracy of direction estimation with massive multiple-input multiple-output (MIMO) receive array in \cite{LiYifanf-machine-learning}.

In recent years, DOA estimation for large-scale MIMO systems has also attracted increasing attention due to its ultra-high angle resolution and precision. But, as the number of antennas tends to large-scale, its complexity and circuit cost become prohibitive in practical applications\cite{T.Engin-09, Handbook-19}. To address this issue, in \cite{8290952}, a low-complexity high-accuracy hybrid analog-digital (HAD) structure method was proposed, but it requires about $M-1$ time slots to infer the true direction angle, where $M$ is the number of antennas per subarray. Therefore, a fast ambiguous DOA elimination method of finding the true emitter direction using only two-time-slot was designed in \cite{chen2021fast}.
In \cite{zhang2021direction}, the authors have formulated the DOA estimation problem of merging three hybrid structures including fully-connected, sub-connected, and switches-based hybrid into a unified framework, with the compression matrix
in a time-varying form. \par
To reduce the high computational complexity and address the failure to fully utilize structural information caused by MIMO systems, in \cite{8400482}, a novel deep learning based super-resolution DOA estimator was proposed. Furthermore, in \cite{8845653}, the authors proposed a low-complexity deep-learning estimator for a MIMO system with a uniform circular array (UCA) at the base station to solve the problem of only providing estimates of source bearings relative to the array axis in a typical uniform linear array (ULA). In \cite{9570330}, DOA estimation was performed for non-circular sources using a large uniform linear array of single snapshots, and the computational complexity was reduced in the process by using a Newton-Raphson iterative method.
In \cite{SBH-ADC}, the authors made an analysis of performance loss on DOA estimation using a MIMO receive array with low-resolution analog-to-digital convertors (ADCs) and found the fact that three bits is sufficient to achieve a trivial performance loss.\par


%
In general, the eigen-decomposition-based DOA measurement methods of using the fully-digital (FD) MIMO receiver have the computational complexity order $O(N^3)$ float-point operations (FLOPs), where $N$ is the number of antennas. As the number of antennas tends to large-scale or ultra-large-scale, the complexity will grow dramatically, for example, when $N=1024$, the complexity will reach up to $ 10^{9} $ FLOPs. Apparently, this complexity is prohibitive for practical applications. To address this issue, in this paper, we develop low-complexity estimators. The main contributions of our work are summarized as follows.\par

\begin{enumerate}
  \item To reduce the computational complexity as much as possible, a low complexity partitioned subarray auto-correlation combining (PSAC) estimator is proposed, where the total array is divided into $K$ subarrays. Here, the covariance matrix of each subarray is independently computed to output the associated estimated  DOA , and then $K$ estimated DOA values are  coherently weighted to achieve the optimal DOA estimation. Compared to the case without partition, the proposed PSAC  makes a significant complexity reduction at the cost of a substantial performance loss.
  \item To improve the accuracy of the PSAC method, a low complexity partitioned subarray cross-correlation combining (PSCC) estimator is proposed. After calculating the cross-correlation matrix (CCM) among sub-arrays, the relationship between CCM and the auto-covariance matrix (ACM) can be utilized to generate multiple candidates for the true direction. The above PSAC is adopted to output a coarse DOA estimation to eliminate the pseudo-solutions, and finally $K(K-1)/2$ optimal DOA estimated values are coherently combined. The proposed PSCC makes a significant complexity reduction while achieving the Cramer-Rao lower bound (CRLB).
  \item To further reduce the complexity and keep a accuracy of direction finding, a new framework of power iteration max correlation successive convex approximation (PI-Max-CSCA) estimator via the continuous convex approximation technique is proposed. The estimating framework consists of three steps. Considering its iterative property, in the first step, a high-performance estimator employs only a small part of the total large array to provide a good initial values of DOA and array manifold vector (AMV). In the second step, based on the initial AMV, the power iteration (PI) method is presented to find the more precise AMV of the emitter. Eventually, using the more precise AMV, the maximum correlation rule via SCA is proposed to estimate the final DOA. Simulations results show that the proposed PSCC and PI-Max-CSCA can achieve the corresponding CRLB of FD MIMO receiver with a much lower complexity. As $N$ tends to large-scale or ultra-large-scale, the proposed methods are one-magnitude to three-magnitude lower in complexity than the FD MIMO with no partition.
\end{enumerate}
The remainder of this paper is organized as follows. Section II describes the system model of fully-digital large-scale array. In Section III, three estimators are proposed, and their performance and computational complexities are also analyzed. We present our simulation results in Section IV. Finally, we draw conclusions in Section V.\par
$Notations$: throughout this paper, boldface lower case and upper case letters represent vectors and matrices, respectively. Signs $(\cdot)^*$, $(\cdot)^H$, $(\cdot)^{-1}$, \textbf{Tr}$(\cdot)$, and $\|\cdot\|$ denote the conjugate operation, conjugate transpose operation, inverse operation, trace operation, and 2-norm operation, respectively. The notation $\textbf{I}_M$ is the $M\times M$ identity matrix. The sign $\mathbb{E}\{\cdot\}$ represents the expectation operation, $\textbf{diag}(\cdot)$ denotes the diagonal operator, $arg(\cdot)$ means the argument of a complex number.

\section{System model}
The emitter signal impinges on the uniformly-spaced linear array (ULA) with $N$ antenna elements, and different antenna element receives the different delayed versions of the same signal.
In the presence of $Q$ emitters impinging from the direction $\theta= [\theta_1,\cdots,\theta_Q]$, the receive signal vector at array is given by
\begin{align}\label{y}
\mathbf{y}=\mathbf{A}\mathbf{s}+\mathbf{w}
\end{align}
where $ \mathbf{A}=[\mathbf{a}(\theta_1),\cdots,\mathbf{a}(\theta_Q)] $ is the array manifold, $\mathbf{s}=[s_1,\cdots,s_Q]^T$ is the Q emitter signals, $\mathbf{w}\sim\mathcal{C}\mathcal{N}(0,\sigma^2_w\textbf{I}_N)$ is the additive white Gaussian noise (AWGN) vector, and $\mathbf{a}(\theta_q)$ is defined as
$
\textbf{a}(\theta_q)=[1,e^{j\frac{2\pi d\sin\theta_q}{\lambda}},\cdots,e^{j\frac{2\pi(N-1)d\sin\theta_q}{\lambda}}]^T,
$
where $\lambda$ is the wavelength of the carrier frequency, and $d=\frac{\lambda}{2}$. Here, the phase reference point is chosen to be the left of the array.
\section{Proposed three low-complexity structures and estimators}
In this section, to reduce the high complexities of conventional DOA measurement methods like Root-MUSIC \cite{1993The} as the number of antennas approaches large-scale, three low-complexity DOA estimators are proposed and the CRLB is derived for the PSAC structure. Finally, the complexity and its CRLB analysis are presented.
\subsection{Proposed PSAC}
\begin{figure}[h]
\centering
\includegraphics[width=0.33\textwidth]{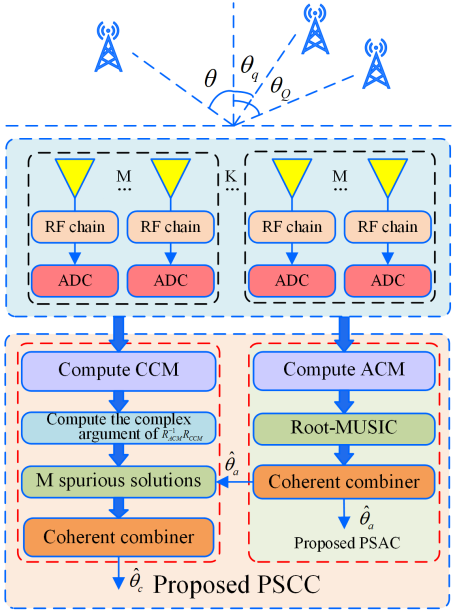}\\
\caption{Proposed a low-complexity  partitioned  structure (Cross-covariance matrix (CCM), Auto-covariance matrix (ACM))}\label{PSAC1.eps}
\end{figure}
The right part of Fig.~\ref{PSAC1.eps} shows the structure of proposed PSAC estimator. The basic idea of the PSAC structure is to divide the total $N$ antennas into $ K $ subarrays with each subarray containing $M$ antennas with $N=KM$. Then, the received $M$ dimensional vector $\textbf{y}_k$  of subarray $k$ can be written as
\begin{align}
\textbf{y}_k(t)=\textbf{a}_k(\theta_q)s_{k}(t)+\textbf{w}_k(t),k=1,2,...,K
\end{align}
where $\textbf{a}_k(\theta_q)$ is the array manifold of the kth subarray, defined as
$
\textbf{a}_k(\theta_q)=[e^{j2\pi\frac{(k-1)Md\sin\theta_q}{\lambda}},\cdots,e^{j2\pi\frac{(kM-1)d\sin\theta_q}{\lambda}}]^T,
$ with sample covariance matrix and the corresponding eigen-decomposition given by
\begin{align}\label{eg}
\textbf{R}_{PS,k}=\frac{1}{L}\sum^L_{l=1}\textbf{y}_k(l)\textbf{y}^H_k(l)=\textbf{U}\Sigma\textbf{U}^H=[\textbf{U}_S\,\textbf{U}_N]\Sigma[\textbf{U}_S\,\textbf{U}_N]^H
\end{align}
where  matrices $ \textbf{U}_S $  and $ \textbf{U}_N $ stand for the signal and noise subspaces, respectively. Based on the above eigenvalue decomposition (EVD), we have the MUISC method as follows
\begin{align}\label{rd}
\overline{\theta}_{k}=\mathop{\arg\max}\limits_{\theta} \frac{1}{\|\textbf{U}^H_N\textbf{a}_k(\theta)\|^2}
\end{align}
which may be computed via the Root-MUSIC, and the corresponding estimated angle is denoted as  $\overline{\theta}_{k}$. Due to the limit on paper length, the detailed process of Root-MUSIC is omitted here\cite{1993The}.

To improve the DOA estimate precision, coherently combining all $K$ sub-array  DOA estimate outputs yields the optimal output as
\begin{align}\label{cc}
\hat{\theta}_a=\frac{1}{K}\sum^K_{k=1}\overline{\theta}_{k},k=1,2,...,K
\end{align}

the CRLB of coherent combiner is derived as
\begin{align}
\text{var}(\hat{\theta})=\frac{\text{CRLB}_k}{K}\approx\frac{\lambda^2}{8\pi^2KLSNR\cos^2\theta_q\overline{d^2}}
\end{align}

This completes the derivation and analysis of the proposed PSAC. Although the PSAC method has an extremely low complexity, there exists a $10\log_{10}K$ dB performance loss compared to the case without partition due to the fact only autocorrelation per subarray is utilized. Therefore, the cross-correlation among  sub-arrays is further exploited to achieve the near-CRLB  performance with the help of auto-correlation.
\subsection{Proposed PSCC}
Fig.~\ref{PSAC1.eps} shows the proposed partitioned array structure. The $N$ antennas are divided into $K$ sub-arrays with each sub-array containing $M$ antennas. $K_s=K(K-1)/2$ CCMs are first computed to find the relationship between the CCM and the ACM, and the coarse DOA estimation $\hat{\theta}_a$ is given by $K$ ACMs, which can be used to eliminate the spurious solutions. The final DOA estimation $\hat{\theta}$ is derived by coherently combining the $K_s$ DOA estimation. 


Any two subarray output vectors $\textbf{y}_k(t)$ and $\textbf{y}_{k+i}(t)$ are used to compute the CCM $\textbf{R}_{k,k+i}$, and
the relationship between the ACM and the CCM is found as
\begin{align}
\textbf{R}_{k,k+i}=e^{-j\frac{2\pi}{\lambda}iMd\sin\theta}\textbf{R}_{ACM,k}^{\frac{1}{2}}(\textbf{R}_{ACM,k+i}^{\frac{1}{2}})^H
\end{align}

Let us define
\begin{align}
\textbf{Z}=\textbf{R}_{k,k+i}^{-1}\textbf{R}_{ACM}=e^{j\frac{2\pi}{\lambda}iMd\sin\theta}\textbf{E}
\end{align}
where $\textbf{R}_{ACM}=\textbf{R}_{ACM,k}^{\frac{1}{2}}(\textbf{R}_{ACM,k+i}^{\frac{1}{2}})^H$, and $\textbf{E}$ is the identity matrix. Taking the trace of $\textbf{Z}$ as $z$, the $M$ solutions are calculated as
\begin{align}
\theta_{k,j}=\arcsin\left(\frac{\lambda(\arg z+2\pi j)}{2\pi iMd}\right),j\in\{0,1,\cdots,M-1\}.
\end{align}
$\hat{\theta}_a$ in (\ref{cc}) is then used to eliminate $M-1$ pseudo-solutions
\begin{align}
\theta_{k_s}=\mathop{\arg\min}_{\theta_{k,j}} | \theta_{k,j}-\hat{\theta}_a|
\end{align}

Coherently combine the $\theta_{k_s}$ generated by $K_s=\frac{K(K-1)}{2}$ CCM to obtain the optimal DOA estimation
\begin{align}
\hat{\theta}_{c}=\frac{1}{K_s}\sum^{K_s}_{k_s=1}\theta_{k_s},k_s=1,2,...,K_s.
\end{align}

\subsection{Proposed PI-Max-CSCA}
\begin{figure}[h]
\centering
\includegraphics[width=0.28\textwidth]{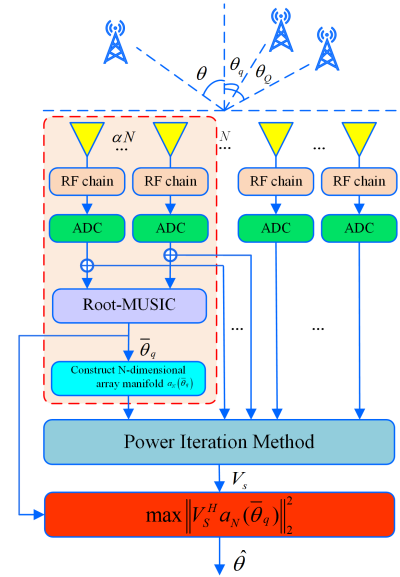}\\
\caption{Proposed a low-complexity PI-Max-CSCA structure}\label{midiedai.eps}
\end{figure}

To further reduce complexity and give a different high-performance solution, the structure of the proposed PI-Max-CSCA method with $N$ antennas is shown in Fig. ~\ref{midiedai.eps}. In this structure, in order to obtain the optimal DOA estimation, a sub-array with a small number of $N_0=\alpha N$ from the $N$ antennas is selected to estimate the initial DOA, where $\alpha\in(0 ,1)$ is much less than 1, which can provide an extremely low-complex initial estimate $\theta_q$ and an extended initial array manifold vector $\mathbf{a}(\bar{\theta_q})$. Then, taking $\mathbf{a}(\bar{\theta_q})$ as the initial input vector, the power iterative operation is performed on the covariance matrix $\mathbf{R}$ of all antennas to form a precise steering vector $\mathbf{V }_s$, which is used to make a more accurate DOA estimation by SCA.
In Fig.~2, the receive signal vector $ \textbf{y}_{N_0} $ at subarray with $ N_0 $ antennas of the qth received signal can be represented as
\begin{align}
\textbf{y}_{N_0}(t)=\textbf{a}_{N_0}(\theta_q)s_{N_0}(t)+\textbf{w}_{N_0}(t)
\end{align}

The initial  angle $\bar{\theta}_q$  of emitter is estimated by Root-MUSIC. Based on the estimated DOA, $N$-dimensional array manifold is constructed as $\textbf{a}(\bar{\theta}_q)=[1,e^{j\frac{2\pi d\sin\bar{\theta}_q}{\lambda}},\cdots,e^{j\frac{2\pi(N-1)d\sin\bar{\theta}_q}{\lambda}}]^T$
which is the initial iterative vector $\textbf{x}_0$ of PI method \cite{2016Numerical}.
\begin{align}\label{f}
\textbf{x}_{\mathbf{n}}=\mathbf{R}\textbf{x}_\textbf{{n-1}}=\mathbf{R}^{n}\textbf{x}_0=\mathbf{R}^{n} \mathbf{a}_{N}\left(\bar{\theta}_{q}\right)
\end{align}

It is assumed that the above PI converges at iteration $n$, then $\textbf{V}_s=\textbf{x}_\textbf{{n}}$  is a more precise steering vector estimate, i.e., steering vector, which is adopted to compute the final DOA by the following criterion.
\begin{align}
\hat\theta=\mathop{\arg\max}\limits_{\theta}\left\|\textbf{V}_{s}^H\mathbf{a}_{N}\left(\theta\right)\right\|_{2}^{2}
\end{align}

Let us define the objective function
\begin{align}
J(\theta)=\left\|\textbf{V}_{s}^H\mathbf{a}_{N}\left(\theta\right)\right\|_{2}^{2}
\end{align}

Since $ J(\theta) $ is a nonlinear function of $\theta$, a successive convex approximation  is a good way to find an approximate solution  \cite{2014Successive}. Defining $ \textbf{w} = \mathbf{a}_{N}\left(\theta\right) $, the above objective function is simplified as
\begin{align}
J(\textbf{w},\textbf{w}^*)&=\textbf{w}^H\textbf{V}s\textbf{V}s^H\textbf{w}
\end{align}


Take its first derivative with respect to $\theta$
\begin{align}\label{e}
\frac{dJ(\theta)}{d\theta}=\frac{dJ(\textbf{w},\textbf{w}^*)}{d\theta}=\frac{d\textbf{w}^H}{d\theta}\textbf{V}_s\textbf{V}_s^H\textbf{w}+\textbf{w}^H\textbf{V}_s\textbf{V}_s^H\frac{d\textbf{w}}{d\theta}
\end{align}

Taking the first derivative of the array manifold has the following form
\begin{align}\label{c}
\frac{d\textbf{w}}{d\theta}=\dot{\mathbf{a}}_{N}\left(\theta\right)=\frac{d}{d \theta} \mathbf{a}_{N}\left(\theta\right)=j(\frac{2 \pi}{\lambda}) \cos \theta \textbf{D}\mathbf{a}_{N}\left(\theta\right)
\end{align}
where
\begin{align}
\textbf{D}=\operatorname{diag}\left\{\left(\left[d_{1}, d_{2}, \ldots, d_{N}\right]\right)\right\}
\end{align}

Substituting (\ref{c}) in (\ref{e}) gives
\begin{small}
\begin{align}
\frac{dJ(\theta)}{d\theta}&=j \frac{2 \pi}{\lambda} \cos \theta\textbf{a}_N(\theta)^H \textbf{V}_{s}\textbf{V}_{s}^H\textbf{D}\textbf{a}_N(\theta)\\\nonumber
&-j \frac{2 \pi}{\lambda} \cos \theta \textbf{a}_N(\theta)^H\textbf{D}\textbf{V}_{s}\textbf{V}_{s}^H\textbf{a}_N(\theta)
\end{align}
\end{small}
Furthermore, the second derivative of $ J(\theta) $ is
\begin{small}
\begin{equation}
\begin{aligned}
&\frac{d^{2} J\left(\textbf{w}, \textbf{w}^{*}\right)}{d \theta^{2}} =j \frac{2 \pi}{\lambda} \sin \theta \cos \theta \textbf{w}^{H} \textbf{D} \textbf{V}_{s} \textbf{V}_{s}^{H} \textbf{w}-j \frac{2 \pi}{\lambda} \cos \theta \bullet \\
&\Big (\frac{d\textbf{w}^{H}}{d\theta} \textbf{D} \textbf{V}_{s} \textbf{V}_{s}^{H} \textbf{w}+\textbf{w}^{H} \textbf{D} \textbf{V}_{s} \textbf{V}_{s}^{H} \frac{d\textbf{w}}{d\theta}\Big)-j \frac{2 \pi}{\lambda} \sin \theta \cos \theta \textbf{w}^{H} \textbf{V}_{s} \textbf{V}_{s}^{H} \textbf{D} \textbf{w}\\
&+j \frac{2 \pi}{\lambda} \cos \theta\left(\frac{d\textbf{w}}{d\theta}^{H} \textbf{V}_{s} \textbf{V}_{s}^{H} \textbf{D} \textbf{w}+\textbf{w}^{H} \textbf{V}_{s} \textbf{V}_{s}^{H} \textbf{D} \frac{d\textbf{w}}{d\theta}\right)
\end{aligned}
\end{equation}
\end{small}
which is further simplified as
\begin{small}
\begin{equation}
\begin{aligned}
&\frac{d^{2} J\left(\theta\right)}{d\theta^{2}}=j\frac{2\pi}{\lambda} \sin\theta \cos\theta \textbf{a}_N(\theta)^H\textbf{D}\textbf{V}_s\textbf{V}_s^H\textbf{a}_N(\theta)-j\frac{2\pi}{\lambda} \sin\theta \cos\theta \bullet\\
&\textbf{a}_N(\theta)^H\textbf{V}_s\textbf{V}_s^H\textbf{D}\textbf{a}_N(\theta)+\frac{8 \pi^{2}}{\lambda^{2}} \cos ^{2} \theta \textbf{a}_{N}(\theta)^{H}\textbf{D}\textbf{V}_{s}\textbf{V}_{s}^{H}\textbf{D} \textbf{a}_{N}(\theta) \\
&- \frac{4 \pi^{2}}{\lambda^{2}} \cos ^{2} \theta \textbf{a}_{N}(\theta)^{H}\textbf{D}^{2}\textbf{V}_{s}\textbf{V}_{s}^{H} \textbf{a}_{N}(\theta)\\
&- \frac{4 \pi^{2}}{\lambda^{2}} \cos ^{2} \theta \textbf{a}_{N}(\theta)^{H}\textbf{V}_{s}\textbf{V}_{s}^{H}\textbf{D}^{2} \textbf{a}_{N}(\theta)
\end{aligned}
\end{equation}
\end{small}

For convenience, let us define,
\begin{small}
\begin{equation}
\begin{aligned}
&A(\theta)=\textbf{a}_{N}(\theta)^{H} \textbf{V}_{s} \textbf{V}_{s}^{H} \textbf{D} \textbf{a}_{N}(\theta) &B(\theta)=\textbf{a}_{N}(\theta)^{H} \textbf{D} \textbf{V}_{s} \textbf{V}_{s}^{H} \textbf{a}_{N}(\theta)\\
&C(\theta)=\textbf{a}_{N}(\theta)^{H} \textbf{D} \textbf{V}_{s} \textbf{V}_{s}^{H} \textbf{D} \textbf{a}_{N}(\theta)
&D(\theta)=\textbf{a}_{N}(\theta)^{H} \textbf{D}^{2} \textbf{V}_{s} \textbf{V}_{s}^{H} \textbf{a}_{N}(\theta)\\
&E(\theta)=\textbf{a}_{N}(\theta)^{H} \textbf{V}_{s} \textbf{V}_{s}^{H} \textbf{D}^{2} \textbf{a}_{N}(\theta)\\
\end{aligned}
\end{equation}
\end{small}

Eventually,  the original $ J(\theta) $  is quadratically approximated as
\begin{small}
\begin{align}
&J(\theta_n+\Delta\theta_n)\approx J(\theta_n)+J'(\theta_n)\Delta\theta_n+\frac{1}{2!}J''(\theta_n)\Delta\theta_n^2\nonumber\\
&=\textbf{a}_{N}\left(\theta_{n}\right)^H\textbf{V}_{s}\textbf{V}_{s}^{H}\mathbf{a}_{N}\left(\theta_{n}\right)
+j \frac{2 \pi}{\lambda} \cos \theta_n\Big(A(\theta_n)-B(\theta_n)\Big) \Delta\theta_n\nonumber\\
&+\Big( j \frac{2 \pi}{\lambda} \sin \theta_n\cos\theta_n\left(B(\theta_n)-A(\theta_n)\right)+\frac{8\pi^2}{\lambda^2}\cos(\theta)^2C(\theta_n)\nonumber\\
&-\frac{4\pi^2}{\lambda^2}\cos(\theta)^2\left(D(\theta_n)+E(\theta_n)\right)\Big) \Delta\theta_n^{2}
\end{align}
\end{small}

Take the first derivative with respect to $ \Delta\theta_n $
\begin{align}
0=J^{\prime}\left(\theta_{n}\right)+J^{\prime \prime}\left(\theta_{n}\right) \Delta \theta_n
\end{align}

so,
\begin{align}
&\Delta \theta_n =-J^{\prime}\left(\theta_{n}\right)(J^{\prime \prime}\left(\theta_{n}\right))^{-1} \\\nonumber
\end{align}

Using the above approximation, the $n$th iterative angle is given by
\begin{align}\label{r}
\theta_{n+1}=\theta_n+\Delta\theta_n
\end{align}

define the  error between iterations $n$ and $n-1$
\begin{align}
P_n=|J(\theta_n) - J(\theta_{n-1})|
\end{align}
which is used to terminate the iteration process.

\subsection{Complexity Analysis}
Below, we make an analysis of computational complexities of the proposed three estimators with fully-digital traditional Root-MUSIC algorithm as a complexity reference. Thus, the complexity of PSAC is as follows
$C_{PSAC}=O\{K\left(M^3-M^2+ML(2M+1)\right)\}$ FLOPs. The complexity of PSCC is $C_{PSCC}=O\{M^3-M^2+ML(2M+1)+\frac{K(K-1)}{2}M^3\}$ FLOPs. The complexity of PI-Max-CSCA is $C_{PMC}=O\{N_0^3-N_0^2+N_0L(2N_0+1)+NL(1+2N)+(\beta-1)N^2\}$ FLOPs. Considering $N$ is far larger than $N_0$, $M$, $\alpha$, $K$ and $L$, compared with the FD Root-MUSIC estimator, the computational complexity of the proposed two estimators is significantly reduced, especially as the number of antennas tends to large-scale.

\section{Simulation Results}
\begin{figure*}[ht]
 \setlength{\abovecaptionskip}{-5pt}
 \setlength{\belowcaptionskip}{-10pt}
 \centering
 \begin{minipage}[t]{0.33\linewidth}
  \centering
  \includegraphics[width=2.5in]{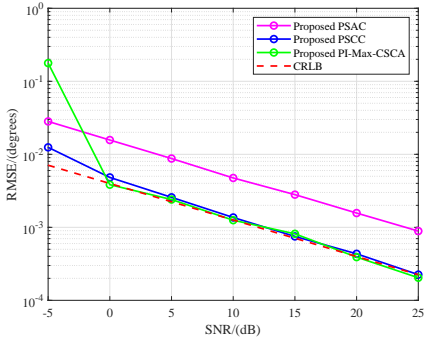}
  \caption{RMSE versus SNR of the proposed method}
  \label{snrrmse1.eps}
 \end{minipage}%
 \begin{minipage}[t]{0.33\linewidth}
  \centering
  \includegraphics[width=2.5in]{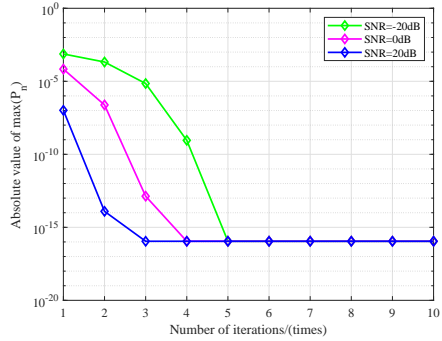}
  \caption{Absolute value of max versus number of iterations}
  \label{fig4}
 \end{minipage}
 \begin{minipage}[t]{0.33\linewidth}
  \centering
  \includegraphics[width=2.5in]{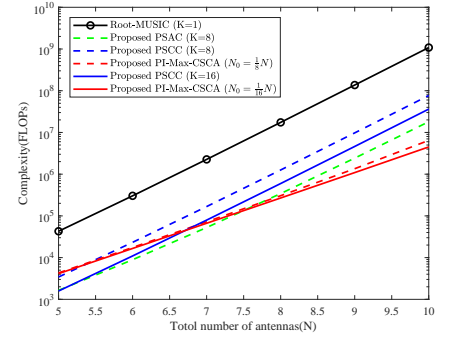}
  \caption{Complexity versus $ \log_2N $($N$ is the number of antennas) } 
  \label{complex2.eps}
 \end{minipage}
\end{figure*}
In this section, we present simulation results to assess the performance of the three DOA estimators: PSAC, PSCC and PI-Max-CSCA with FD Root-MUSIC as a performance benchmark. A source impinge on the array and $L = 1$. In massive/ultra-massive MIMO scenarios, the number $N$ of antennas at receive array varies from 32 to 1024.\par

Fig. \ref{snrrmse1.eps} plots the root mean squared error (RMSE) versus SNR of the three proposed DOA estimators PSAC, PSCC and PI-Max-CSCA for $ N=1024 $, and $ M=N0=256 $, where the corresponding CRLB is used as a performance benchmark. From Fig. 3, it is seen that the proposed PSCC and PI-Max-CSCA methods can achieve the corresponding CRLB as snapshot is 1 and the performance of the PSAC method can close to the corresponding CRLB as snapshot is 1 \par

%

To show the impact of SNR on the number of iterations of SCA in our proposed PI-Max-CSCA method, Fig. 4 demonstrates the absolute value of max $ P_n $ versus the number of iterations of the proposed method as SNR ranges from -20dB to 20dB, given a fixed $ N=1024 $, and $ M=N_0=256 $. From this figure, it is seen that the convergence speed becomes faster and more stable as the SNR increases. The number of iterations are 3,~4,~and 5 for three distinct SNRs -20dB, 0dB, and 20dB, respectively. \par

Fig. \ref{complex2.eps} shows the computational complexities versus the number of antennas with $M=N_0\in\{64,128\}$ and $ N $ varying from 32 to 1024. From this figure, it is seen that as the number of total antennas increases, the complexities of all methods increase gradually. However, as the number of antennas goes to ultra-large-scale, the computational complexities of our proposed three methods are an-order-of-magnitude to three-order-of-magnitude lower in terms of FLOPs than conventional Root-MUSIC.
\section{Conclusions}
In this paper, based on the large-scale/ultra-large-scale MIMO receive array, Three low-complexity DOA estimators are proposed: PSAC, PSCC and PI-Max-CSCA. As the number of antennas tends to large-scale or ultra-large-scale, they make a significant complexity reduction compared with conventional Root-MUSIC. The PSAC is one or two-order magnitude less in complexity than Root-MUISC at the expense of a substantial precision loss. The PSCC and PI-Max-CSCA can achieve the CRLB with one to three-order-magnitude reduction in complexity.
Using these two methods makes DOA estimation for massive/ultra massive MIMO receivers feasible for future practical applications such as 5G-evolution.

\ifCLASSOPTIONcaptionsoff
  \newpage
\fi

\bibliographystyle{IEEEtran}
\bibliography{IEEEfull,reference}
\end{document}